\documentclass[prd,aps,showpacs,nofootinbib]{revtex4}
\usepackage{amsmath}
\usepackage{amsfonts}
\usepackage{amssymb}
\usepackage[dvips]{graphicx}
\usepackage{color}

\begin{document}

\title{Lorentz-violating contributions of the Carroll-Field-Jackiw model to
the CMB anisotropy}
\author{Rodolfo Casana}
\thanks{casana@ufma.br}
\author{Manoel M. Ferreira Jr.}
\thanks{manojr07@ibest.com.br}
\author{Josberg S. Rodrigues}
\thanks{josberg@ufma.br}
\affiliation{Departamento de F\'{\i}sica, Universidade Federal do Maranh\~{a}o (UFMA),
Campus Universit\'{a}rio do Bacanga, S\~{a}o Lu\'{\i}s-MA, 65085-580 - Brasil}

\begin{abstract}
We study the finite temperature properties of the
Maxwell-Carroll-Field-Jackiw (MCFJ) electrodynamics for a purely
spacelike background. Starting from the associated finite
temperature partition function, a modified black body spectral
distribution is obtained. We thus show that, if the CMB radiation is
described by this model, the spectrum presents an anisotropic
angular energy density distribution. We show, at leading order, that
the Lorentz-breaking contributions for the Planck's radiation law
and for the Stefan-Boltzmann's law are nonlinear in frequency and
quadratic in temperature, respectively. Using our results, we set up
bounds for the Lorentz-breaking parameter, and show that Lorentz
violation in the context of the MCFJ model is unable to yield the
known CMB anisotropy (of 1 part in $10^{5})$.
\end{abstract}

\maketitle

\section{Introduction}

In his seminal work \cite{Maxwell}, Maxwell proposed that light
requires a medium to travel, in a full analogy to the experiences
involving waves propagation in fluids. Such medium was named as
ether. In view of the already observed light properties, it was
assumed that the ether permeated the whole space, was of a
negligible density, and had imperceptible interaction with matter.
However, the ether was abandoned with the advent of the special
theory of relativity \cite{Einstein} that established Lorentz
covariance as one of the fundamental symmetries of nature. Nowadays,
the Lorentz covariance pervades all the field theories describing
fundamental interactions and has the status of a cornerstone in the
construction of all modern physical theories. The present
experiments confirm Lorentz invariance to a very high precision at
currently accessible energy scales that goes up to 2 TeV. The new
experiments to be performed in the Large Hadron Collider (LHC) at
CERN, that will extend the energy scale to approximately 14 TeV,
should test the Lorentz symmetry to confirm that it still remains
unspoiled or to reveal some indications about its violation. At the
moment, the possibility is discussed of Lorentz and CPT symmetry
breaking at Planck scale (or in the very early Universe when
energies are close to the Planck scale). One such scenario is
suggested by string theory \cite{string} and it is a key feature of
noncommutative field theories \cite{nocom}.

The researches about Lorentz and CPT violation are commonly performed under
the framework of the standard model extension (SME) developed by Colladay
and Kostelecky \cite{Kostelecky}. The SME is an enlarged version of the
usual standard model that embraces all Lorentz-violating coefficients
(generated as vacuum expectation values of tensor quantities belonging to a
fundamental theory defined at Planck scale) that yield Lorentz scalars (as
tensor contractions) in the observer frame. Such coefficients rule Lorentz
violation in the particle frame, where they are seen as sets of independent
numbers, whereas they work out as genuine tensor in the observer frame. A
strong motivation to study the SME is the necessity to get some information
about underlying physics to the Planck scale where the Lorentz symmetry may
be broken due to quantum gravity effects. The photon sector of the SME has
been extensively studied with a double purpose: the determination of new
electromagnetic effects induced by the Lorentz-violating (LV) interactions
and the imposition of stringent upper bounds for the magnitudes of the LV
coefficients. Lorentz violation has been investigated in a broad perspective
in the latest years \cite{SME,Ted}.

The research about LV effects on classical electromagnetism was started by
Carroll-Field-Jackiw \cite{CFJ}, who studied the Maxwell electrodynamics in
the presence of the assigned Carroll-Field-Jackiw (CFJ) term $\epsilon ^{\mu
\nu \kappa \lambda }\left( k_{AF}\right) _{\mu }A_{\nu }F_{\kappa \lambda }$%
, with $\left( k_{AF}\right) _{\mu }$ standing for the LV fixed background.
The gauge sector of the SME embodies a CPT-odd CFJ\ term and the CPT-even
one, $W^{\mu \nu \kappa \lambda }F_{\mu \nu }F_{\kappa \lambda },$ both of
which imply vacuum birefringence \cite{CFJ,pol1,pol2}, which takes place
whenever the light velocity depends on the polarization mode, amounting to a
rotation in the polarization plane. While the CFJ term yields a causal,
stable, and unitary electrodynamics only for a purely spacelike background
\cite{Adam}, the CPT-even term provides an electrodynamics not plagued with
stability illness. Regarding that birefringence increases linearly with the
distance traveled, the analysis of this effect over cosmological scales
offers an exceptionally sensitive signal for Lorentz violations. Within this
context, as the cosmic microwave background (CMB) is partially polarized, it
can be considered an experimental optical probe able to catch minuscule
Lorentz violations \cite{cmb1,cmb2}. In \cite{cmb2}, the spacelike sector of
the CFJ background, $k_{AF}$, has been analyzed and has been shown that
experimental data from the Boomerang experiment and the 5-year Wilkinson
Microwave Anisotropy Probe (WMAP) survey are consistent with weak Lorentz
violation at the 1-sigma level. Other issues concerning LV corrections to
the Planckian spectrum were addressed in Ref. \cite{winder}, where the
emission and absorption of radiation by nonrelativistic electrons in the SME
framework was properly regarded to derive such effects.

The CMB, which is the oldest thermal radiation available to observation,
constitutes a good scenario to be described by the Lorentz-violating photon
sector of the SME at finite temperature. The detected CMB is interpreted as
compelling evidence for the big bang, since theoretical nucleosynthesis
calculations foresees the existence of a cosmic background radiation at a
temperature of some kelvins \cite{Gamov}. The data coming from the Cosmic
Background Explorer (COBE) and WMAP revealed that the CMB is a perfect
Planckian black body distribution at 2.73 K with high precision of one part
in $10^{5}$, which bounds the anisotropies to this small extent \cite%
{Cobe,Wmap}.

Considering that the light propagation is affected by the Lorentz
violation, it is probable that its thermodynamical properties and
spectral distribution are also altered. Therefore the black body
pattern of the CMB is an interesting phenomenon where
Lorentz-violating effects may play a relevant role, mainly when
concerned with the anisotropies of the spectrum, an issue that
captures broad attention nowadays
\cite{Cobe,Wmap,quadrupole,hofmann}. A natural framework to deal
with black body radiation and Lorentz violation is the finite
temperature field theory \cite{Kapusta}. The aim of the present work
is to study the finite temperature properties of the MCFJ
electrodynamics for the case of a purely spacelike background (for
which the model provides a positive-definite Hamiltonian). Indeed,
taking as starting point the MCFJ Lagrangian, we can construct the
partition function for this gauge model (after the constraints
structure is well determined). Such partition function provides all
thermodynamical information required, including the energy density
distribution. We thus show that if the CMB radiation is described by
the MCFJ model, the radiation presents an anisotropic angular energy
density distribution. Consequently, it is possible to obtain the
Lorentz-breaking contributions to the Planck's radiation law and to
the Stefan-Boltzmann's law.

This paper is outlined as follows. In Sec. \ref{sec-2}, we develop the
Hamiltonian analysis of the constraints structure of the MCFJ model by
following the Dirac formalism for constrained systems. In Sec. \ref{sec-3},
we construct the partition function (into the functional formalism) and
study the thermodynamical properties of the model, including the modified
energy density distribution and the modified Stefan-Boltzmann's law. In the
last section, we present our conclusions and final remarks.

\section{The MCFJ electrodynamics: Hamiltonian structure \label{sec-2}}

The Maxwell-Carroll-Field-Jackiw model is defined by the following
Lagrangian density:
\begin{equation}
\mathcal{L}=-\frac{1}{4}F_{\mu \nu }F^{\mu \nu }-\frac{1}{4}\epsilon ^{\mu
\nu \kappa \lambda }\left( k_{AF}\right) _{\mu }A_{\nu }F_{\kappa \lambda },
\end{equation}%
where $F_{\mu \nu }=\partial _{\mu }A_{\nu }-\partial _{\nu }A_{\mu
}$ is the electromagnetic stress tensor, $(k_{AF})_{\mu}$ is the
Lorentz-breaking vector background, and $\epsilon ^{\mu \nu \kappa
\lambda }$ is the totally antisymmetric Levi-Civita tensor with
$\epsilon ^{0123}=1$. The corresponding Euler-Lagrange equation for
the vector field is
\begin{equation}
\partial _{\nu }F^{\nu \mu }+\left( k_{AF}\right) _{\nu }\tilde{F}^{\mu \nu
}=0,  \label{mm-7}
\end{equation}%
where $\tilde{F}^{\nu \alpha }=\displaystyle\frac{1}{2}\epsilon ^{\nu \alpha
\mu \beta }F_{\mu \beta }$ is the dual tensor. Since the pioneering work of
Carroll-Field-Jackiw \cite{CFJ}, the properties of the MCFJ electrodynamics
were extensively investigated in several distinct respects \cite%
{Adam,Cerenkov, Casana}. In the present work, the goal is to evaluate the LV
corrections to the Planck black body distribution, which will be done by
means of the imaginary-time formalism for finite temperature field theory.
Once the partition function is carried out, the entire thermodynamics of the
model becomes available. For it, we should first try to understand the
constraint structure of the model, unveiled by a careful Hamiltonian
analysis.

In order to accomplish the Hamiltonian analysis of this model, we begin
defining the canonical conjugate momentum
\begin{equation}
\pi ^{\mu }=-F^{0\mu }-\frac{1}{2}\epsilon ^{0\mu \alpha \beta }\left(
k_{AF}\right) _{\alpha }A_{\beta },  \label{mm8}
\end{equation}%
with which we can write the fundamental Poisson brackets (PB): $\displaystyle%
\left\{ A_{\mu }\left( x\right) ,\pi ^{\nu }\left( y\right) \right\} =\delta
_{\mu }^{\nu }\delta \left( \mathbf{x-y}\right) $.

From Eq.(\ref{mm8}), it is easy to note that $\pi ^{0}=0$; such a null
momentum yields a primary constraint $\phi _{1}=\pi ^{0}\approx 0$ (into the
Dirac formalism, the symbol $\approx $ denotes a \textit{weak equality}).
Also, the momenta $\pi ^{k}$ are\ defined via the following dynamic
relation:
\begin{equation}
\pi ^{k}=\dot{A}_{k}-\partial _{k}A_{0}-\frac{1}{2}\epsilon ^{0kij}\left(
k_{AF}\right) _{i}A_{j},  \label{mm-9}
\end{equation}%
while the canonical Hamiltonian density is explicitly written as
\begin{eqnarray}
{\mathcal{H}_{C}} &=&\frac{1}{2}\left( \pi ^{k}\right) ^{2}+\pi
^{k}\partial
_{k}A_{0}+\frac{1}{4}\left( F_{jk}\right) ^{2}  \notag \\
&&+\frac{1}{2}\pi ^{k}\epsilon ^{0kij}\left( k_{AF}\right) _{i}A_{j}+\frac{1%
}{8}\left[ \epsilon ^{0kij}\left( k_{AF}\right) _{i}A_{j}\right]
^{2}\notag\\
&&+\frac{1}{4}\epsilon ^{0kij}\left( k_{AF}\right) _{0}A_{k}F_{ij}-\frac{1}{4%
}\epsilon ^{0kij}\left( k_{AF}\right) _{k}A_{0}F_{ij}. \label{mm-7a}
\end{eqnarray}%
Following the usual Dirac procedure, we introduce the primary Hamiltonian $%
\left( H_{P}\right) $ by adding to the canonical Hamiltonian all the primary
constraints, $H_{P}=H_{C}+\displaystyle\int \!\!d^{3}\mathbf{y~}C\pi ^{0}$,
where $C$ is a bosonic Lagrange multiplier. The consistency condition of the
primary constraint, $\dot{\pi}^{0}=\left\{ \pi ^{0},H_{P}\right\} \approx 0$%
, gives a secondary constraint
\begin{equation}
\phi _{2}=\ \partial _{k}\pi ^{k}\ +\frac{1}{4}\epsilon ^{0kij}\left(
k_{AF}\right) _{k}F_{ij}\approx 0,
\end{equation}%
which reveals that the usual Gauss's law is modified by an arbitrary
Lorentz-breaking background. It can be written in terms of the electric and
magnetic fields: $\,\nabla \cdot \mathbf{E}+\left( \mathbf{k}_{AF}\right)
\cdot \mathbf{B}=0$. Therefore, for the case of a pure timelike background,
no modification is implied on the usual Gauss law. The situation changes for
a pure spacelike case, for which the Gauss's law is modified by the presence
of the background. Such modification reflects the coupling between the
electric and magnetic sectors in the MCFJ electrodynamics \cite{CFJ,Casana}.

The consistency condition of the modified Gauss's law gives $\dot{\phi}%
_{2}=\left\{ \phi _{2},H_{P}\right\} =0$. Thus, the secondary constraint is
automatically conserved and there are no more constraints in this model. The
bosonic multiplier of the primary constraint remains undetermined. It is an
evidence of the existence of first-class constraints, such as it can be
verified by computing the PB between the constraints $\left\{ \phi _{1},\phi
_{2}\right\} =0$. Therefore, the set of constraints
\begin{equation}
\phi _{1}=\pi ^{0}\approx 0,~\ \ \ \phi _{2}=\ \partial _{k}\pi ^{k}\ +\frac{%
1}{4}\epsilon ^{0kij}\left( k_{AF}\right) _{k}F_{ij}\approx 0,
\end{equation}%
is a first-class one.

\subsection{Equations of motion and gauge fixing conditions}

Following the Dirac conjecture, we define the extended Hamiltonian $\left(
H_{E}\right) $ by adding all the first-class constraint to the primary
Hamiltonian,%
\begin{equation}
H_{E}=H_{C}+\int d\mathbf{y~}\left[ C\phi _{1}+D\phi _{2}\right] .
\end{equation}%
Under this Hamiltonian, we compute the time evolution of the canonical
variables of the system:
\begin{eqnarray}
\dot{A}_{0} &=&\left\{ A_{0},H_{E}\right\} =C, \\
\dot{A}_{k} &=&\pi ^{k}+\partial _{k}A_{0}+\frac{1}{2}\epsilon ^{0kij}\left(
k_{AF}\right) _{i}A_{j}-\partial _{k}D,  \label{mm-30}
\end{eqnarray}%
showing that the dynamics of $A_{0}$ and $A_{k}$ remain arbitrary. For $%
\pi ^{0}$\ and $\pi ^{k}$, it is attained%
\begin{equation}
\dot{\pi}^{0}=\partial _{k}\pi ^{k}+\frac{1}{4}\epsilon ^{0kij}\left(
k_{AF}\right) _{k}F_{ij}=\phi _{2}\approx 0,
\end{equation}%
\begin{eqnarray}
\dot{\pi}^{k} &=&-\frac{1}{2}\pi ^{l}\epsilon ^{0lik}\left( k_{AF}\right)
_{i}-\partial _{j}F_{kj}  \notag \\
&&-\frac{1}{4}\epsilon ^{0lij}\epsilon ^{0lak}\left( k_{AF}\right)
_{i}\left( k_{AF}\right) _{a}A_{j}-\frac{1}{2}\left( k_{AF}\right)
_{0}\epsilon ^{0kij}F_{ij}  \notag \\
&&-\frac{1}{2}\epsilon ^{0kli}\left( k_{AF}\right) _{l}\partial _{i}A_{0}+%
\frac{1}{2}\epsilon ^{0kli}\left( k_{AF}\right) _{l}\partial _{i}D.
\end{eqnarray}%
Making use of Eq. (\ref{mm-30}), these expressions can be rewritten as%
\begin{eqnarray}
&&\partial _{k}F^{0k}+\frac{1}{4}\epsilon ^{0ikj}\left(
k_{AF}\right)
_{i}F_{k}{}_{j}+\partial _{k}\partial _{k}D =0,  \notag \\[0.2cm]
& &\partial _{\alpha }F^{\alpha k}+\frac{1}{2}\epsilon ^{k\alpha \mu
\nu }\left( k_{AF}\right) _{\alpha }F_{\mu \nu }
\label{ss1}\\[0.15cm]
&&\hspace{1cm}+\epsilon^{0kli}\left( k_{AF}\right) _{l}\partial
_{i}D-\partial _{0}\partial _{k}D =0. \notag
\end{eqnarray}

We can see that Eqs. (\ref{mm-30}) and (\ref{ss1}) are similar to the
Lagrangian equations (\ref{mm-9}) and (\ref{mm-7}), respectively, if and
only if $D=0$. Thus, we should impose a gauge condition in such a way to fix
$D=0$. It is known that the Dirac algorithm requires a number of gauge
conditions equal to the number of first-class constraints in the theory.
However, those gauge conditions must be compatible with the Euler-Lagrange
equations, such that they should fix $D=0$ and determine the Lagrangian
multiplier $C$. Such that the gauge conditions together with the first-class
constraints should form a second-class set.

\subsubsection{Radiation gauge}

Equation (\ref{mm-30}) suggests the following gauge fixing condition $\psi
_{1}=\partial _{k}A_{k}\approx 0,$ whose consistency relation, $\dot{\psi}%
_{1}=\left\{ \psi _{1},H_{E}\right\} \approx 0$, gives
\begin{equation}
\nabla ^{2}A_{0}-\frac{1}{2}\epsilon ^{0kij}\left( k_{AF}\right)
_{k}F_{ij}-\nabla ^{2}D\approx 0.
\end{equation}%
In order to get an equation only for $D$, we impose
\begin{equation}
\psi _{2}=\nabla ^{2}A_{0}-\frac{1}{2}\epsilon ^{0kij}\left( k_{AF}\right)
_{k}F_{ij}\approx 0,
\end{equation}%
as a second gauge condition, such that $\nabla ^{2}D=0$ is used to fix $D=0$%
. The consistency condition of the second gauge condition, $\dot{\psi}%
_{2}=\left\{ \psi _{2},H_{E}\right\} \approx 0$, implies%
\begin{equation}
\nabla ^{2}C-\frac{1}{2}\epsilon ^{0kij}\left( k_{AF}\right) _{k}\dot{F}%
_{ij}=0.
\end{equation}%
Thus, we have determined all the Lagrange multipliers. Therefore, the set $%
\Sigma _{a}=\left\{ \phi _{1}~,~\phi _{2}~,~\psi _{1}~,~\psi _{2}\right\} ~$%
is a second-class one while the corresponding PB matrix, defined as
$ M_{ab}\left( x,y\right) =\left\{ \Sigma _{a}\left( x\right)
,\Sigma _{b}\left( y\right) \right\}$, explicitly reads as a
nonsingular matrix,
\begin{equation}
M\left( x,y\right) =\left(
\begin{array}{cccc}
0 & 0 & 0 & -\nabla ^{2} \\[0.2cm]
0 & 0 & \nabla ^{2} & 0 \\[0.2cm]
0 & -\nabla ^{2} & 0 & 0 \\[0.2cm]
\nabla ^{2} & 0 & 0 & 0%
\end{array}%
\right) \delta \left( \mathbf{x-y}\right) ,  \label{mm-43}
\end{equation}%
whose inverse
\begin{equation}
M^{-1}\left( x,y\right)=\left(
\begin{array}{cccc}
0 & 0 & 0 & -G\left( \mathbf{x-y}\right) \\[0.2cm]
0 & 0 & G\left( \mathbf{x-y}\right) & 0 \\[0.2cm]
0 & -G\left( \mathbf{x-y}\right) & 0 & 0 \\[0.2cm]
G\left( \mathbf{x-y}\right) & 0 & 0 & 0%
\end{array}%
\right) ,  \label{mm-48}
\end{equation}%
is written in terms of the Green function for the Poisson equation, $G\left(
\mathbf{x-y}\right) $, given by
\begin{eqnarray}
\nabla ^{2}G\left( \mathbf{x-y}\right) &=&-\delta \left( \mathbf{x-y}\right)
,~\ \   \notag \\
&& \\
\ \ G\left( \mathbf{x}\right) &=&\int \frac{d\mathbf{p}}{\left( 2\pi \right)
^{3}}\frac{~e^{i\mathbf{p\cdot x}}}{\mathbf{p}^{2}}=\frac{1}{4\pi \left\Vert
\mathbf{x}\right\Vert }.  \notag
\end{eqnarray}

At this level, it is necessary to assure that the set of first-class
constraints and gauge fixing conditions become strong equalities. Such
requirement is fulfilled by defining a new bracket operation, the Dirac
brackets, $\left\{ \cdot ,\cdot \right\} _{D}$, as
\begin{eqnarray}
\left\{ A\left( x\right) ,B\left( y\right) \right\} _{D} &=&\left\{ A\left(
x\right) ,B\left( y\right) \right\} -\int \!\!d\mathbf{u}\!\!\int \!\!d%
\mathbf{v~}\left\{ A\left( x\right) ,\Sigma _{c}\left( u\right) \right\}
\notag \\
&&\times \left[ M^{-1}\left( u,v\right) \right] _{cd}\left\{ \Sigma
_{d}\left( v\right) ,B\left( y\right) \right\} ,
\end{eqnarray}%
where $\Sigma _{a}=\left\{ \phi _{1},~\phi _{2},~\psi _{1},~\psi
_{2}\right\} $ and $M^{-1}\left( x,y\right) $ is the inverse matrix defined
in Eq. (\ref{mm-48}).

Thus, the non-null Dirac brackets for the physical variables
\begin{eqnarray}
\left\{ A_{k}\left( x\right) ,\pi _{j}\left( y\right) \right\} _{D} &=&-
\left[ \delta _{kj}-\frac{\partial _{k}\partial _{j}}{\nabla ^{2}}\right]
\delta \left( \mathbf{x-y}\right) ,  \label{dblv1} \\
\left\{ \pi ^{k}\left( x\right) ,\pi ^{j}\left( y\right) \right\} _{D} &=&%
\frac{1}{2}\epsilon ^{0jli}\left( k_{AF}\right) _{l}\partial
_{i}^{x}\partial _{k}^{x}G\left( \mathbf{x-y}\right)  \notag \\
&&-\frac{1}{2}\epsilon ^{0kli}\left( k_{AF}\right) _{l}\partial
_{i}^{x}\partial _{j}^{x}G\left( \mathbf{x-y}\right) ,  \label{dblv2} \\
\left\{ A_{0}\left( x\right) ,\pi ^{k}\left( y\right) \right\} _{D} &=&%
\mathbf{-}\epsilon ^{0kli}\left( k_{AF}\right) _{l}\partial _{i}^{x}G\left(
\mathbf{x-y}\right) ,  \label{dblv3}
\end{eqnarray}%
should be compared with the algebra of the pure Maxwell electrodynamics (in
the radiation gauge), for which the only non-null Dirac bracket is given by
Eq. (\ref{dblv1}). Also, the MCFJ algebra establishes a noncommutative
relation for the transverse momenta which can be contrasted with the
noncommutative gauge field algebra $\;\left[ A_{k}\left( x\right)
,A_{j}\left( y\right) \right] =i\mathcal{\ell }_{jk}\delta \left( \mathbf{x-y%
}\right) \ ,\ \left[ A_{k}\left( x\right) ,\pi _{j}\left( y\right) \right]
=i\delta _{kj}\delta \left( \mathbf{x-y}\right) \ ,\ \left[ \pi _{k}\left(
x\right) ,\pi _{j}\left( y\right) \right] =0,$ proposed in Ref. \cite%
{kfields} for studying black body radiation in a Lorentz-breaking context.
In a definitive way, we can infer that the physical properties of the MCFJ
model are very different from the Maxwell theory and from the noncommutative
gauge field approach. Consequently, these models should have different
thermodynamical properties such as will be clearly shown in the last section.

Under the Dirac brackets, the canonical Hamiltonian (\ref{mm-7a}) reads as
\begin{equation}
H=\int \!\!d^{3}\mathbf{y}\left[ \frac{1}{2}\mathbf{E}^{2}+\frac{1}{2}%
\mathbf{B}^{2}+\frac{1}{2}\left( k_{AF}\right) _{0}\mathbf{A}\cdot \mathbf{B}%
\right] .
\end{equation}%
While a pure timelike background does not guarantee a positive-definite
Hamiltonian, it can be obtained for the case of a pure spacelike background,
for which a well-defined quantum theory may be constructed. Indeed, in this
case the model can be quantized in the canonical formalism, once the
canonical commutation relations for the quantum fields are obtained from the
Dirac brackets (by means of the correspondence principle). It may be also
quantized via the functional integral formalism. We follow the last
quantization procedure to compute the partition function and to analyze the
thermodynamical properties of the MCFJ model.

\section{The partition function \label{sec-3}}

In this section, we study the thermodynamical properties of the MCFJ model.
The fundamental object for this analysis is the partition function. The
Hamiltonian analysis performed in the previous section allows one to define
in a correct way the functional integral representation of the partition
function, which is given by
\begin{eqnarray}
Z\left( \beta \right) &=&\int \!\!\mathcal{D}A_{\mu }\mathcal{D}\pi ^{\mu
}\delta \left( \phi _{1}\right) \delta \left( \phi _{2}\right) \delta \left(
\psi _{1}\right)  \notag \\
&&\times \delta \left( \psi _{2}\right) ~\left\vert \det \left\{ \Sigma
_{a}\left( x\right) ,\Sigma _{b}\left( y\right) \right\} \right\vert ^{1/2}
\notag \\
&&\times \exp \left\{ \int_{\beta }\!\!dx~\left( i\pi ^{\mu }\partial _{\tau
}A_{\mu }-\mathcal{H}_{C}\right) \right\} ,
\end{eqnarray}%
where $\Sigma _{a}=\left\{ \phi _{1},~\phi _{2},~\psi _{1},~\psi
_{2}\right\} $ is a second-class set formed by the first-class constraints
and the gauge fixing conditions, $M_{ab}\left( x,y\right) =\left\{ \Sigma
_{a}\left( x\right) ,\Sigma _{b}\left( y\right) \right\} $ is the constraint
matrix given in Eq. (\ref{mm-43}), whose determinant is $\det M\left(
x,y\right) =\det \left( -\nabla ^{2}\right) ^{4}$. Given the bosonic
character of the gauge field, its functional integration can be performed
over all the fields satisfying periodic boundary conditions in the $\tau \ $%
variable: $A\left( \tau ,\mathbf{x}\right) =A\left( \tau +\beta ,\mathbf{x}%
\right) $. The short notation $\displaystyle\int_{\beta }dx$ denotes\textbf{%
\ }$\displaystyle\int_{0}^{\beta }\!\!d\tau \!\!\int \!\!d^{3}\mathbf{x,}$
and $\mathcal{H}_{C}$ is the canonical Hamiltonian given by Eq. (\ref{mm-7a}%
).

We first compute the integration on the field $\pi ^{0}$. Using a Fourier
representation for $\delta \left( \phi _{2}\right) ,$
\begin{eqnarray}
\ \delta \left( \phi _{2}\right) &=&\int \mathcal{D}\Lambda ~\exp \left\{
i\int_{\beta }\!\!dx~\Lambda \left[ \frac{{}}{{}}\partial _{k}\pi ^{k}\
\right. \right.  \notag \\
&&~\ \ \ \ \ \ \ \ \ \ \left. \left. +\frac{1}{4}\epsilon ^{0kij}\left(
k_{AF}\right) _{k}F_{ij}\right] \right\} ,
\end{eqnarray}%
doing the change $\Lambda \rightarrow \Lambda +iA_{0},$ and performing the
integration over the $\pi ^{k}$ field, the partition function reads as
\begin{eqnarray}
Z\left( \beta \right) &=&\det \left( -\nabla ^{2}\right) ^{2}\int \mathcal{D}%
A_{\mu }\mathcal{D}\Lambda ~\delta \left( \psi _{1}\right) \delta \left(
\psi _{2}\right) ~  \notag \\
&&\times \exp \left\{ \int_{\beta }\!\!dx~-\frac{1}{2}\left( \partial _{\tau
}A_{k}-\partial _{k}\Lambda \right) ^{2}\right.  \notag \\
&&\left. -\frac{i}{2}\left( \partial _{\tau }A_{k}-\partial _{k}\Lambda
\right) \epsilon ^{0kij}\left( k_{AF}\right) _{i}A_{j}\right\} \   \notag \\
&&\times \exp \left\{ \int_{\beta }\!\!dx~\frac{i}{4}\Lambda \epsilon
^{0kij}\left( k_{AF}\right) _{k}F_{ij}\right.  \notag \\
&&\left. -\frac{1}{4}\left( F_{jk}\right) ^{2}-\frac{1}{4}\epsilon
^{0kij}\left( k_{AF}\right) _{0}A_{k}F_{ij}\right\} .
\end{eqnarray}%
The integration of the $A_{0}$ field gives the contribution $\left[ \det
\left( -\nabla ^{2}\right) \right] ^{-1}$. At once, we rename $\Lambda
=A_{\tau }$ and $\left( k_{AF}\right) _{0}=i\left( k_{AF}\right) _{\tau }$,
and setting $\epsilon ^{0kij}=-i\epsilon _{\tau kij}~,~\epsilon _{\tau
123}=1,$ we get the partition function for the MCFJ model in the Coulomb
gauge:
\begin{eqnarray}
Z\left( \beta \right) &=&N\det \left( -\nabla ^{2}\right) \int \mathcal{D}%
A_{a}\mathcal{~}\delta \left( \partial _{k}A_{k}\right)  \notag \\
&&\hspace{-1cm}\times \exp \left\{ \int_{\beta }\!\!dx~-\frac{1}{4}%
F_{ab}F_{ab}-\frac{1}{4}\,\epsilon _{abcd}\left( k_{AF}\right)
_{a}A_{b}F_{cd}\right\} ,
\end{eqnarray}%
where $a,b,c,d=\tau ,1,2,3$.

The partition function in the Coulomb gauge is not explicitly covariant. It
is well known that if the covariance is explicit, the calculation process
becomes more manageable. The procedure to pass from a noncovariant gauge to
a covariant one, like the Lorentz gauge $\partial _{a}A_{a}=0$, can be
implemented using the Faddeev-Popov ansatz, which is defined by
\begin{equation}
\int D\omega (x)\,\delta \left( G\left[ A_{a}^{\omega }\right] \right)
\,\det \left\vert \frac{\delta G\left[ A_{a}^{\omega }\right] }{\delta
\omega }\right\vert _{\omega =0}\equiv 1,  \label{fot-39}
\end{equation}%
where $\omega \left( x\right) $ is the gauge parameter, $D\omega $ is a
gauge group measure, $G\left[ A_{a}\right] $ is a covariant gauge fixing
condition, $A_{a}^{\omega }$ is the gauge-transformed field $\left(
A_{a}^{\omega }=A_{a}+\partial _{a}\omega \right) ,$ and $\displaystyle\det
\left\vert \frac{\delta G\left[ A_{a}^{\omega }\right] }{\delta \omega }%
\right\vert _{\omega =0}$ is the so-called Faddeev-Popov determinant, which
is gauge invariant. We thus choose the Lorentz gauge
\begin{equation}
G\left[ A_{a}\right] =-\frac{1}{\sqrt{\xi }}\partial _{a}A_{a}+f~,
\end{equation}%
$f$ being an arbitrary scalar function and $\xi $ a gauge parameter. In this
way, we have $G\left[ A_{a}^{\omega }\right] =G\left[ A_{a}\right] -\square
\omega /\sqrt{\xi }$, which implies
\begin{equation}
\det \left\vert \frac{\delta G\left[ A_{a}^{\omega }\right] }{\delta \omega }%
\right\vert _{\omega =0}=\det \left\vert \frac{-\square }{\sqrt{\xi }}%
\right\vert ,
\end{equation}%
where $\square =\partial _{a}\partial _{a}=\left( \partial _{\tau }\right)
^{2}+\nabla ^{2}$. As the partition function is independent of $f$, such a
factor can be eliminated by integrating it with the weight $\displaystyle%
\exp \left( -\frac{1}{2}\int_{\beta }dx~f^{2}\right) $. Thus, after an
integration by parts, the partition function takes the form:
\begin{eqnarray}
Z\left( \beta \right) &=&\int D{A}_{a}~\det \left\vert \frac{-\square }{%
\sqrt{\xi }}\right\vert \exp \left\{ \int_{\beta }dx-\frac{1}{2}A_{a}\left[
\frac{{}}{{}}-\square \delta _{ab}\right. \right.  \notag  \label{eq-33} \\
&&\left. \left. -\left( \frac{1}{\xi }-1\right) \partial _{a}\partial
_{b}-S_{ab}\right] A_{b}\right\} .
\end{eqnarray}%
where have defined the operator $S_{ab}=\epsilon _{acdb}\left( k_{AF}\right)
_{c}\partial _{d}$. For convenience, we choose the Feynman gauge $\xi =1$,
and the integration over the gauge field gives
\begin{equation}
Z\left( \beta \right) =\det \left( -\square \right) ~\left[ \det \left(
-\square \delta _{ab}-S_{ab}\right) \right] ^{-1/2}.  \label{mm-142a}
\end{equation}%
After some algebra, we obtain $\det \left( -\square \delta
_{ab}-S_{ab}\right) =\left[ \det \left( -\square \right) \right] ^{2}\det
\left( \square ^{2}+\left( k_{AF}\right) ^{2}\square -\left( \left(
k_{AF}\right) \cdot \partial \right) ^{2}\right) $. Replacing it in the
partition function (\ref{mm-142a}), we obtain
\begin{equation}
Z\left( \beta \right) =Z_{A}\left( \beta \right) Z_{LV}\left( \beta \right) .
\end{equation}%
Here, the quantity $Z_{A}\left( \beta \right) =\exp \left\{ -\text{tr}\ln
\left( -\square \right) \right\} $ is the partition function of the usual
electromagnetic field (without Lorentz violation), while
\begin{equation}
Z_{LV}\left( \beta \right) =\displaystyle\exp \left\{ -\frac{1}{2}\text{Tr}%
\ln \left[ 1+\frac{\left( k_{AF}\right) ^{2}}{\square }-\frac{\left( \left(
k_{AF}\right) \cdot \partial \right) ^{2}}{\square ^{2}}\right] \right\} ,
\label{zlv}
\end{equation}%
is the contribution stemming from the Chern-Simon-like LV term. We can
compute the involved trace writing the gauge field in terms of a Fourier
expansion,
\begin{equation}
A_{a}(\tau ,\mathbf{x})=\left( \frac{\beta }{V}\right) ^{\frac{1}{2}}\sum_{n,%
\mathbf{p}}e^{i(\omega _{n}\tau +\mathbf{x}\cdot \mathbf{p})}\tilde{A}_{a}(n,%
\mathbf{p}),
\end{equation}%
where $V$ represents the system volume and $\omega _{n}$ are the bosonic
Matsubara's frequencies, $\omega _{n}=\displaystyle\frac{2n\pi }{\beta }~$,
for $n=0,1,2,\cdots $.

\subsection{The pure electromagnetic contribution}

We start computing the pure electromagnetic contribution,%
\begin{equation}
\ln Z_{A}\left( \beta \right) =-\text{tr}\ln \left( -\square \right)
=-\sum\limits_{n,\mathbf{p}}\ln \left( \beta ^{2}\left[ \mathbf{p}%
^{2}+\left( \omega _{n}\right) ^{2}\right] \right) .
\end{equation}

The sum in $n$ is evaluated as
\begin{equation}
\sum_{n=-\infty }^{+\infty }\ln \left[ (2\pi n)^{2}+\beta ^{2}\omega _{%
\mathbf{p}}^{2}\right] =\beta \omega _{\mathbf{p}}+2\ln \left[ 1-e^{-\beta
\omega _{\mathbf{p}}}\right] ,  \label{soma-1}
\end{equation}%
where $\omega _{\mathbf{p}}=\left\Vert \mathbf{p}\right\Vert $. Therefore,
the contribution of the pure electromagnetic field is%
\begin{equation}
\ln Z_{A}\left( \beta \right) =-2V\int \frac{d^{3}\mathbf{p}}{(2\pi )^{3}}%
\left[ \frac{\beta \omega _{\mathbf{p}}}{2}+\ln (1-e^{-\beta \omega _{%
\mathbf{p}}})\right] ,
\end{equation}%
where the sum in $\left\Vert \mathbf{p}\right\Vert $\ was replaced by an
integral. The latter integral can be explicitly evaluated in spherical
coordinates, $\mathbf{p\equiv }\left( \omega ,\theta ,\phi \right) $, where $%
\omega =$ $\omega _{\mathbf{p}}=\left\Vert \mathbf{p}\right\Vert $, $\theta $
is the angle between the background $\mathbf{k}_{AF}$ and the photonic
momentum $\mathbf{p}$, while $\phi $ is the azimuthal angle. In this way,
the partition function reads as
\begin{eqnarray}
\ln Z_{A}\left( \beta \right) &=&-\frac{2V}{\left( 2\pi \right) ^{3}}\int
d\Omega \int_{0}^{\infty }d\omega ~\left[ \frac{\beta \omega ^{3}}{2}\right.
\notag \\
&&\left. \frac{{}}{{}}+\omega ^{2}\ln (1-e^{-\beta \omega })\right] ,
\label{planck-0}
\end{eqnarray}%
with $d\Omega =\sin \theta d\theta d\phi $ being the differential
solid-angle element. Neglecting the vacuum contributions, the partition
function is exactly carried out
\begin{equation}
\ln Z_{A}\left( \beta \right) =V\frac{\pi ^{2}}{45\beta ^{3}}.
\label{planck-0a}
\end{equation}

The energy density $\left( u_{A}=U_{A}/V\right)$ for the pure
electromagnetic field which represents the expectation value of the energy
per unit volume (over the thermodynamical ensemble) can be easily obtained $%
\left( u_{A}=-V^{-1}\partial \ln Z_{A}/\partial \beta \right) $, yielding:
\begin{equation}
u_{A}=\int_{0}^{\infty }d\omega ~\frac{1}{\pi ^{2}}\frac{\omega ^{3}}{%
e^{\beta \omega }-1}.  \label{planck-0b}
\end{equation}%
Without integrating in frequency, we obtain the energy density of radiation\
per frequency unity, that is, the well-known Planck distribution for the
black body spectrum:
\begin{equation}
u_{A}(\omega )=\frac{1}{\pi ^{2}}\frac{\omega ^{3}}{e^{\beta \omega }-1}.
\label{dp-1}
\end{equation}

Now, performing the integral in frequency in Eq. (\ref{planck-0b}), we get
the total energy density in the cavity
\begin{equation}
u_{A}=\frac{\pi ^{2}}{15\beta ^{4}}=aT^{4},  \label{ua-1}
\end{equation}%
which corresponds to the Stefan-Boltzmann law, while the constant $a$%
\footnote{%
In SI unities the relation is $\sigma =\displaystyle\frac{ac}{4}$ with $c$
the vacuum light velocity and $a=\displaystyle\frac{\pi ^{2}k_{B}^{4}}{
15(\hbar c)^{3}}=7.565604554\times 10^{-16}~$Jm$^{-\text{3}}$K$^{-\text{4}}\
,\ \sigma =5.670277968\times 10^{-8}~$Wm$^{-\text{2}}$K$^{-\text{4}}$, where
$k_{B}$ is the Boltzmann's constant and $\hbar $ the Planck's constant.} is
related to the Stefan-Boltzmann's constant $\left( \sigma \right) $ by $%
a=4\sigma $. The energy density per solid-angle element is
\begin{equation}
u_{A}\left( \beta ,\Omega \right) d\Omega =\frac{\pi }{60}\frac{1}{\beta ^{4}%
}d\Omega ,  \label{ua-2}
\end{equation}%
which stands for a perfect isotropic distribution.

\subsection{The CPT-odd and Lorentz-violating contribution}

The Lorentz-breaking contribution, $Z_{LV}\left( \beta \right) $, to the
partition function is computed for the case of a pure spacelike background $%
k_{AF}=\left( 0,\mathbf{k}_{AF}\right) $, once it is known that the
Hamiltonian is positive-definite (stable) only for this background
configuration. Such a feature guarantees the existence of the functional
integral from which is attained the partition function, thus, the partition
function (\ref{zlv}) reads
\begin{equation}
\ln Z_{LV}=-\frac{1}{2}\sum\limits_{n,\mathbf{p}}\ln \left( 1-\frac{\mathbf{k%
}_{AF}^{2}}{\left( \omega _{n}\right) ^{2}+\mathbf{p}^{2}}+\frac{\left(
\mathbf{k}_{AF}\cdot \mathbf{p}\right) ^{2}}{\left[ \left( \omega
_{n}\right) ^{2}+\mathbf{p}^{2}\right] ^{2}}\right) .  \label{zlv2}
\end{equation}%
Now, we consider the spacelike background as a weak coupling, $\left\Vert
\mathbf{k}_{AF}\right\Vert \ll 1$, then we get at order $\mathbf{k}_{AF}^{2}$
\begin{equation}
\ln Z_{LV}=\frac{1}{2}\sum\limits_{n,\mathbf{p}}\left( \frac{\mathbf{k}%
_{AF}^{2}}{\left( \omega _{n}\right) ^{2}+\mathbf{p}^{2}}-\frac{\left(
\mathbf{k}_{AF}\cdot \mathbf{p}\right) ^{2}}{\left[ \left( \omega
_{n}\right) ^{2}+\mathbf{p}^{2}\right] ^{2}}\right)
\end{equation}%
The series above can be computed easily by using expression (\ref{soma-1}).
Performing the sum and expressing the resultant integrals in spherical
coordinates, the partition function takes the form
\begin{eqnarray}
\ln Z_{LV} &=&\frac{1}{2}\frac{\mathbf{k}_{AF}^{2}}{\left( 2\pi \right) ^{3}}%
V\int d\Omega \int_{0}^{\infty }d\omega \left\{ \frac{\beta \omega }{2}+%
\frac{\beta \omega }{e^{\beta \omega }-1}\right\}  \notag \\
&&-\frac{1}{2}\frac{\mathbf{k}_{AF}^{2}}{\left( 2\pi \right) ^{3}}V\int
d\Omega \cos ^{2}\theta \int_{0}^{\infty }d\omega \left\{ \frac{\beta \omega
}{2}+\frac{\beta \omega }{2}\right.  \notag \\
&&\left. \times \frac{1}{e^{\beta \omega }-1}+\frac{\left( \beta \omega
\right) ^{2}}{2}\frac{~e^{\beta \omega }}{~\left( e^{\beta \omega }-1\right)
^{2}}\right\} .  \label{zv-1}
\end{eqnarray}%
Here, the integrals in the frequency $\left( \omega \right) $ can be
performed exactly, implying
\begin{equation}
\ln Z_{LV}=V\frac{\mathbf{k}_{AF}^{2}}{96\pi \beta }\int d\Omega -V\frac{%
\mathbf{k}_{AF}^{2}}{64\pi \beta }\int d\Omega \cos ^{2}\theta \ .
\label{zv-2}
\end{equation}%
Performing now the angular integrations, we obtain the Lorentz-violating
contribution to the partition function,
\begin{equation}
\ln Z_{LV}=V\frac{\mathbf{k}_{AF}^{2}}{48\beta },  \label{planck-1a}
\end{equation}%
where we have neglected vacuum contributions.

From Eq. (\ref{zv-1}), the expectation value of the energy is achieved by
unit volume (over the thermodynamical ensemble) for the Lorentz-breaking
contribution:
\begin{eqnarray}
u_{LV} &=&-\frac{\mathbf{k}_{AF}^{2}}{4}\int_{0}^{\infty }d\omega \frac{1}{%
\pi ^{2}}\frac{\omega ^{3}}{e^{\beta \omega }-1}~\left\{ \frac{5}{6}\frac{1}{%
\omega ^{2}}-\frac{7}{6}\frac{\beta }{\omega }\right.  \notag \\
&&\left. \times \frac{e^{\beta \omega }}{\left( e^{\beta \omega }-1\right) }+%
\frac{1}{6}\beta ^{2}\frac{e^{\beta \omega }\left( e^{\beta \omega
}+1\right) }{\left( e^{\beta \omega }-1\right) ^{2}}\right\} .
\label{planck-2a}
\end{eqnarray}%
The integrand gives the LV corrections to the Planckian energy density
distribution to be
\begin{eqnarray}
u_{LV}\left( \omega \right) &=&-\frac{\mathbf{k}_{AF}^{2}}{4}\frac{1}{\pi
^{2}}\frac{\omega ^{3}}{e^{\beta \omega }-1}~\left\{ \frac{5}{6}\frac{1}{%
\omega ^{2}}-\frac{7}{6}\frac{\beta }{\omega }\frac{e^{\beta \omega }}{%
\left( e^{\beta \omega }-1\right) }\right.  \notag \\
&&\left. +\frac{1}{6}\beta ^{2}\frac{e^{\beta \omega }\left( e^{\beta \omega
}+1\right) }{\left( e^{\beta \omega }-1\right) ^{2}}\right\} ,  \label{dp-2}
\end{eqnarray}%
where nonlinear contributions in the frequency $\omega $ appear.

The Lorentz-breaking contribution to the Stefan-Boltzmann law can be
achieved by integrating Eq. (\ref{planck-2a}), which yields
\begin{equation}
u_{LV}=\frac{\mathbf{k}_{AF}^{2}}{48\beta ^{2}}=\frac{\mathbf{k}_{AF}^{2}}{48%
}T^{2}.  \label{uv-1}
\end{equation}

From Eq. (\ref{zv-2}), we can also determine the LV contribution to the
energy density in each solid angle $\left( d\Omega =\sin \theta d\theta
d\phi \right) $:
\begin{equation}
u_{LV}\left( \beta ,\Omega \right) d\Omega =\frac{\mathbf{k}_{AF}^{2}}{96\pi
\beta ^{2}}\left[ 1-\frac{3}{2}\cos ^{2}\theta \right] d\Omega ~,
\label{uv-2}
\end{equation}%
in which it is manifest the presence of the anisotropy factor $\left( \cos
^{2}\theta \right) $. This reveals that LV is a mechanism that can play an
important role in explaining CMB anisotropies.

\subsection{The MCFJ thermodynamics}

The energy density of the MCFJ\ model is the one associated with the full
partition function (\ref{mm-142a}), given by the sum of the contributions (%
\ref{ua-1}) and (\ref{uv-1}), namely,
\begin{equation}
u_{MCFJ}=aT^{4}+\frac{\mathbf{k}_{AF}^{2}}{48}T^{2}.  \label{S_B}
\end{equation}%
The expression above shows a LV correction to the Stefan-Boltzmann law at $%
\mathbf{k}_{AF}^{2}$ order with a dependence in the temperature as $T^{2}$.
Such a correction is potentially more significant at low temperatures. The
result (\ref{S_B}) can be rewritten in two ways. The first one is
\begin{equation}
u_{MCFJ}=\bar{a}\left( T\right) T^{4},
\end{equation}%
with $\bar{a}\left( T\right) $ being an effective coefficient that retains
the temperature and LV modifications:%
\begin{equation}
\bar{a}\left( T\right) =a+\frac{\mathbf{k}_{AF}^{2}}{48T^{2}}.
\label{bound-1}
\end{equation}%
It affords an opportunity to establish a first bound for the spacelike LV
background by using the experimental data for the Stefan-Boltzmann constant
\cite{codata} $\sigma =\left( 5.67040\pm 0.00004\right) \times 10^{-8}~{%
\text{W}}{\ \text{m}^{\text{-2}}\text{K}^{\text{-4}}}$. Thus, considering
that the corrections to the Stefan-Boltzmann law are of the order of the
experimental error, we get $\left\Vert \mathbf{k}_{AF}\right\Vert \leq
3.6\times 10^{-15}$ GeV for $T=2.73$ K (the black body temperature of the
CMB radiation).

The second form to express Eq. (\ref{S_B}) is by considering the $a$
constant as fixed and attributing the small variations on the energy density
to temperature fluctuations $\left( \delta T\right) $. The expression (\ref%
{S_B}) can be then written as
\begin{equation}
u_{MCFJ}\approx a\left( T^{4}+4T^{3}\delta T\right) ,  \label{bound-2}
\end{equation}%
which gives the temperature fluctuation $\delta T$ with respect to the black
body temperature $T$ without LV interactions. In others words, we assume
that the Stefan-Boltzmann law $u\propto T^{4}$ remains valid for both models
\cite{hofmann}. Considering Eq. (\ref{bound-2}), we write the first order
temperature corrections $\left( \delta T\right) $ as
\begin{equation}
\frac{\delta T}{T}=\frac{u_{MCFJ}-u_{M}}{4u_{M}}.  \label{DT}
\end{equation}%
Such expression allows one to extract the temperature offsets from the MCFJ
(model following the method developed in Ref. \cite{hofmann}), which can be
compared with the experimental data coming from the Far-InfraRed Absolute
Spectrophotometer (FIRAS) and the WMAP. Thus, we see that Lorentz violation
is a mechanism that can play an important role in explaining CMB
anisotropies. Physically, the term $\delta T $ in Eq. (\ref{bound-2}) stands
for the temperature offsets of the MCFJ integrated spectra (integration over
all frequencies) compared to the conventional Maxwell black body integrated
spectra $\left( u_{M}\right) $.

The expression for $\delta T$ leads to a second (but similar) bound for the
LV parameter if we compare the quadrupole fluctuation implied by Lorentz
violation [see Eq. (\ref{uv-2})] with the quadrupole temperature fluctuation
of the CMB \cite{Cobe,Wmap,quadrupole}: $\delta T/T\sim 10^{-6}$, with $%
T=2.73$ K. In this case, we obtain $\left\Vert \mathbf{k}_{AF}\right\Vert
\sim 2.75\times 10^{-15}$ GeV.

Also, from Eqs. (\ref{dp-1}) and (\ref{dp-2}) we derive the energy density
of the radiation per frequency for the MCFJ electrodynamics
\begin{eqnarray}
u_{MCFJ}\left( \omega \right)  &=&\frac{1}{\pi ^{2}}\frac{\omega ^{3}}{%
e^{\beta \omega }-1}\left\{ 1-\frac{\mathbf{k}_{AF}^{2}}{4}\left[ \frac{5}{6}%
\frac{1}{\omega ^{2}}-\frac{7}{6}\frac{\beta }{\omega }\right. \right.
\notag \\
&&\left. \left. \times \frac{e^{\beta \omega }}{\left( e^{\beta \omega
}-1\right) }+\frac{1}{6}\beta ^{2}\frac{e^{\beta \omega }\left( e^{\beta
\omega }+1\right) }{\left( e^{\beta \omega }-1\right) ^{2}}\right] \right\} .
\end{eqnarray}

From the angular energy distribution expressions (\ref{ua-2}) and (\ref{uv-2}%
), we write the MCFJ energy density per solid-angle element:%
\begin{equation}
u_{MCFJ}\left( \beta ,\Omega \right) \,d\Omega =\left[ \frac{\pi }{60}\frac{1%
}{\beta ^{4}}+\frac{\mathbf{k}_{AF}^{2}}{96\pi \beta ^{2}}\left( 1-\frac{3}{2%
}\cos ^{2}\theta \right) \right] d\Omega .
\end{equation}%
Thus, the angular energy distribution at $\mathbf{k}_{AF}^{2}$ order
provides a quadrupole $\left( l=2\right) $ contribution to the power angular
spectrum, revealing an interesting feature: the LV contribution to the
spectrum is anisotropic and gives a maximal contribution in the plane
perpendicular to background direction. From Eq. (\ref{zlv}), it is easy to
show that a contribution at $\left( \mathbf{k}_{AF}^{2}\right) ^{n}$ order
to the power angular spectrum of the black body radiation may be considered.
It gives contributions until the order $l=2n$ at the same time it is
associated with a $T^{4-2n}$ temperature dependence. This guarantees that,
for high temperatures, the relevant contributions stem only from the first
terms of the expansion.

\section{Conclusions and remarks}

In this work, we have initially established the constraint structure
of the MCFJ electrodynamics having as the main goal the correct
evaluation of the partition function of this model (at the finite
temperature regime). With the partition function, the thermodynamics
properties of the model were determined, revealing the LV
corrections to the black body spectral distribution. As the CMB map
is in fact nothing more than a black body radiation pattern only
slightly perturbed by fluctuations, our purpose involves an attempt
of relating the CMB anisotropies with the LV corrections here
evaluated. Indeed, we have addressed what is expected to appear as
anisotropies in the CMB map if the photonic sector in the lately
universe is described by the finite temperature CFJ electrodynamics.
Such calculation shows that the LV CFJ term modifies (in leading
order) the monopole and quadrupole moments of angular power spectrum
in a proper way. This is ascribed to the form of the MCFJ field
algebra, that is different from the Maxwell and from the
noncommutative gauge field (and space-time) approaches proposed in
Refs. \cite{kfields,k-minkowski}. Such difference leads to very
distinct black body spectra. In fact, at order $n\geq 1$ the
temperature corrections to the integrated spectra are proportional
to $T^{4+4n}$, for the noncommutative space-time approach,
$T^{4+2n}$, for the noncommutative gauge field model approach, and
$T^{4-2n}$, for the finite temperature MCFJ model.

Finally, we have seen that the CFJ term is able to induce anisotropic
contributions to the CMB. Although, we should mention that the background
magnitude for yielding a CMB anisotropy of 1 part in $10^{5}$ is
approximately $10^{-15}$ GeV. Considering that birefringence data constrain
such background as tightly as $10^{-33}$ eV, we conclude that Lorentz
violation, as set up in the MCFJ model, cannot be used to explain such
anisotropies.

In a forthcoming work, we intend to present the thermodynamical
contributions for the black body radiation provided by the less constrained
coefficients of the CPT-even term of the gauge sector of the SME. Such work
is in progress.

\begin{acknowledgments}
R. C. thanks Conselho Nacional de Desenvolvimento Cient\'{\i}fico e Tecnol%
\'{o}gico (CNPq) for partial support. M. M. F. is grateful to CNPq and to
FAPEMA (Funda\c{c}\~{a}o de Amparo \`{a} Pesquisa do Estado do Maranh\~{a}o)
for partial support. J. S. R. thanks FAPEMA for full support.
\end{acknowledgments}

\end{document}